\documentclass[reprint,amsmath,amssymb]{revtex4-2}
\usepackage[rgb]{xcolor}
\usepackage{graphicx}
\usepackage{hyperref}
\usepackage{physics}
\usepackage{comment}
\usepackage{enumitem}
\usepackage{hyperref}
\usepackage[capitalize]{cleveref}

\begin{document}

\title{Tunable glassy dynamics in models of dense cellular tissue}

\author{Helen S. Ansell}
\affiliation{Department of Physics, Emory University, Atlanta, GA, USA}
\author{Chengling Li}
\affiliation{Department of Physics, Emory University, Atlanta, GA, USA}
\author{Daniel M. Sussman}
\email{daniel.m.sussman@emory.edu}
\affiliation{Department of Physics, Emory University, Atlanta, GA, USA}

\begin{abstract} 
Observations of glassy dynamics in experiments on confluent cellular tissue have inspired a wealth of computational and theoretical research to model their emergent collective behavior.
Initial studies of the physical properties of several geometric cell models, including vertex-type models, have highlighted anomalous sub-Arrhenius, or ``ultra-strong,'' scaling of the dynamics with temperature.
Here we show that the dynamics and material properties of the 2d Voronoi model deviate even further from the standard glassforming paradigm.
By varying the characteristic shape index $p_0$, we demonstrate that the system properties can be tuned between displaying expected glassforming behavior, including the breakdown of the Stokes-Einstein-Sutherland relation and the formation of dynamical heterogeneities, and an unusual regime in which the viscosity does not diverge as the characteristic relaxation time increase and dynamical heterogeneities are strongly suppressed.
Our results provide further insight into the fundamental properties of this class of anomalous glassy materials, and provide a step towards designing materials with predetermined glassy dynamics.  
\end{abstract}

\maketitle

Collective cell motion within densely packed tissue is a fundamental biological process that requires the coordination of large groups of cells.
Such motion is essential for biological processes including morphogenesis, wound healing and cancer cell migration~\cite{Shaw2009,Ilina2009,Weijer2009}.
Cellular materials inherently exist far from equilibrium and are controlled by processes across a range of length and time scales.
Nevertheless, experiments have shown that confluent cell monolayers have properties reminiscent of those observed in non-living disordered systems~\cite{Angelini2011,Nnetu2012,Park2015,Park2016,Malinverno2017,Atia2018,Vishwakarma2020,Kim2020,Lin2020,Cerbino2021,Sadhukhan2024Review}, and this analogy is used as a common starting point for quantifying their observed dynamics.

These findings have fueled interest in theoretical and computational modeling of these complex collective phenomena, with geometric models such as vertex~\cite{Nagai2001,Farhadifar2007,Bi2015} and Voronoi~\cite{Bi2016} models, and the cellular Potts model~\cite{Graner1992} proving popular choices.
Despite being highly coarse-grained, these models have successfully captured aspects of the glassy dynamics observed in cellular tissue, such as the solid-to-fluid transition as model parameters are varied~\cite{Bi2015,Bi2016,Li2018,Chiang2016,Park2015,Grosser2021}.
Although these models have received substantial attention, many of the properties of these models remain unknown even under purely equilibrium conditions.
In fact, relatively little is known about the extent to which different models of dense tissue are even meaningfully different from each other -- when modeling a real cellular system, do different models simply parameterize the system differently but otherwise capture the same physics, or do these different models display fundamentally different phenomenology? 
We argue that more completely determining the \emph{equilibrium} dynamical and structural properties of these cell models provides a useful baseline for modeling cells in more biologically-relevant conditions. 

From this perspective, the geometric cell models are particularly interesting due to both the predictions they make about the behavior of real cellular materials and their non-standard behavior as coarse-grained statistical-physics models.
Some of the original interest in these models stemmed from their predicted mechanical properties, such as possessing an unusual athermal rigidity transition~\cite{Sussman2018SM,Bi2015,Teomy2018,Pinto2022} which shares interesting features with constraint-satisfaction transitions~\cite{Urbani2023}.
In equilibrium, these models further display a number of unexpected properties. 
Their disordered-phase dynamics display peculiar sub-Arrhenius or ``ultra-strong'' scaling of the alpha-relaxation timescale $\tau_{\alpha}$ with temperature~\cite{Sussman2018EPL,Li2021,Sadhukhan2021} -- a property shared (to our knowledge) only with computational models of low-density vitrimeric polymer glasses~\cite{Ciarella2019}. 
While the detailed many-body nature of the interactions in these systems is surely linked to their anomalous dynamics, understanding both the full set of ways in which these models have properties different from standard glassformers and the fundamental principles that lead to these properties remains an open problem.

To better understand the fundamental phenomenology of these models we examine the Stokes-Einstein-Sutherland (SE) relation~\cite{Einstein1905,Sutherland1905}, a hallmark of glassy materials~\cite{Ediger2000}.
In a standard liquid, the SE ratio $D\eta/T$ between the diffusion coefficient $D$, viscosity $\eta$ and temperature $T$ is expected to remain constant as the temperature of the system is varied.
A characteristic of standard glassforming materials is that this relation is obeyed at high temperatures, but at lower temperatures the onset of glassy dynamics is connected with ``Stokes-Einstein decoupling'' -- the breaking down of this relation~\cite{Rossler1990,Hodgdon1993}.
This is often suggested to result from the formation of dynamical heterogeneities (DH) within the system, with $D$ expected to be dominated by fast-moving populations of particles while $\eta$ is largely determined by the slower-moving particles~\cite{Bordat2003,Sengupta2014,Pastore2015,Parmar2017}.
As such, SE violation is often taken as indirect evidence of the formation of DH within the system.

By studying the SE relation and making direct measurements of DH in the two-dimensional (2d) Voronoi model, we demonstrate that this model is indeed quite different from typical glassforming materials.
We observe that at lower values of the shape index $p_0$, the model displays SE violation and the formation of DH, while at higher $p_0$ values there is no SE breakdown and a corresponding lack of DH. 
We also find an anomalous scaling of the viscosity with temperature, which at higher $p_0$ shows no clear divergence over a range of temperatures in which the characteristic relaxation time changes by over four orders of magnitude. 
Our results provide insight into the atypical fundamental dynamical properties of this model -- adding to the growing evidence that these geometric cell models are in a unique universality class of disordered dynamics.

\begin{figure*}[t]
\centering
\includegraphics[width=0.9\textwidth]{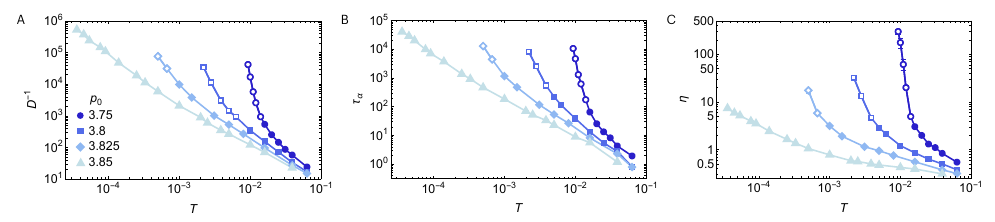}
\caption{Temperature dependence of 
(\textit{A})  the inverse diffusion coefficient $D^{-1}$, 
(\textit{B}) the alpha-relaxation time $\tau_{\alpha}$, and
(\textit{C}) the viscosity $\eta$ of the 2d Voronoi model. 
The color scale in \textit{A} applies to all panels. 
Solid markers indicate points for which the Voronoi model is in the disordered phase at the given $p_0$ and $T$ value while open markers indicate that the system has hexatic ordering. Error bars, which are typically smaller than the marker size, indicate the standard error across ten configurations.
}
\label{fig:T-dependence}
\end{figure*}

\section*{The Voronoi model}
We use a standard implementation of the Voronoi model for $N$ cells in 2d under periodic boundary conditions, for which the dimensionless energy is given by~\cite{Farhadifar2007,Bi2015,Bi2016}
\begin{equation}
e =  \sum_{i=1}^N \left[k_r (a_i - a_{0,i})^2 + (p_i - p_{0,i})^2\right].
\end{equation}
Here, $a_i$ and $p_i$ are the area and perimeter of cell $i$, while $a_{0,i}$ and $p_{0,i}$ are model parameters representing the respective target area and perimeter for that cell. 
The constant $k_r = \frac{l^2 k_A}{k_P}$, where $l$ sets the unit of length and $k_A$ and $k_P$ are the elastic moduli associated with the area and perimeter, is set to one.
We choose the unit of length so that the average dimensionless area of a cell, $a_i = A_i/l^2$, is $\expval{a_i}=1$.
Here we implement a monodisperse version of the model by setting $a_{0,i} = 1$ and $p_{0,i} = p_0$ for all cells.
The shape index $p_0$ characterizes the mechanical stability of the cellular packing at zero temperature. 
In the thermodynamic limit there is a crossover shape index of $p_c \approx 3.81$; for $p_0 \lesssim p_c$ the cells are geometrically frustrated and are mechanically solid ~\cite{Bi2015,Li2018}, and for $p_0 \gtrsim p_c$ the cells are geometrically compatible and mechanically unstable~\cite{Bi2015,Li2018}.
It is in this latter regime that many models of real cell sheets are often parameterized~\cite{Devany2021,Armengol2023}.
Close to the crossover, there is a region of hexatic ordering observed in the model that persists at the boundary between solid-like and fluid-like behavior of the model at finite temperature~\cite{Li2018}.
In the results presented here we indicate state points in the disordered regime of the model with solid markers, while those for which there is hexatic ordering present (as determined from the location of peaks in the susceptibility of the bond orientational order parameter) are shown with open markers.

We investigate the dynamics of this model for different values of the shape index $p_0$ in the canonical (NVT) ensemble.
Temperature is implemented through a Nos\'e-Hoover thermostat to ensure transport properties and thermodynamic distributions are preserved within the system~\cite{Basconi2013}.
We use an integration timestep of $\dd t = 0.01\tau$, where $\tau$ is the dimensionless unit of time.
We set the number of cells $N=4096$ for all results presented in this work.
We have verified that this choice of $N$ is large enough to avoid finite-size effects encountered for smaller system sizes and have verified our results are consistent with those for system sizes up to $N=32\,768$.
All simulations were performed using the open-source cellGPU software~\cite{Sussman2017}.

\section*{Results}

\textbf{Temperature dependence of dynamical quantities.}
Given that the 2d Voronoi model displays unusual sub-Arrhenius scaling of the alpha-relaxation time $\tau_{\alpha}$ with temperature~\cite{Sussman2018EPL,Li2021}, we first examine how other dynamical properties, namely the diffusivity $D$ and viscosity $\eta$, scale with temperature $T$ in the model across a range $p_0$ values that at zero temperature span the range from geometrically frustrated to geometrically compatible.
\Cref{fig:T-dependence}\textit{A} shows the temperature-dependence of $D^{-1}$ (see Methods), which at low $p_0$ values shows a rapid increase over a fairly narrow range of temperatures as the system is cooled, as expected in a glassforming material. 
By contrast, at higher $p_0$ values a similar change in $D^{-1}$ requires several decades of change in $T$. 
This unusual behavior is consistent with that of $\tau_{\alpha}$ (see Methods), shown in \cref{fig:T-dependence}\textit{B}, which highlights that the temperature range studied in each case corresponds to $\tau_{\alpha}$ values up to $\sim 10^4$, which is a typical scale at which to probe the computationally supercooled and glassy regime~\cite{Sussman2018EPL,Kob1995}.
These results highlight that, especially at higher $p_0$ values, the dynamics of the Voronoi model slow down much more slowly than expected in a typical glassformer.

In a typical glassy system the viscosity $\eta$ follows a similar trend to $\tau_{\alpha}$, as would be predicted by standard models of viscoelastic behavior, such as the Maxwell model~\cite{Maxwell1867}.
As such, due to the high computational costs of directly measuring $\eta$ in simulations, it is often assumed that $\tau_{\alpha}\propto\eta$~\cite{Chen2006,Kumar2007,Xu2009,Shi2013} or $\tau_{\alpha}\propto\eta/T$~\cite{Ozawa2016,Ikeda2011PRL,Ikeda2011JChemPhys,Sengupta2013} when examining the SE relation.
These arguments are based on the Maxwell model and the Gaussian solution to the diffusion equation, respectively.
Given that typically in the glassy regime, $\eta$ and $\tau_{\alpha}$ can change by orders of magnitude due to a small changes in $T$, both approximations are usually reasonable when studying SE breakdown~\cite{Shi2013}.
Here we determine $\eta$ directly \emph{via} the VSS RNEMD technique~\cite{Kuang2012} (see Methods).
As we will highlight in the following section, the anomalous scaling of the dynamics means that in this model using $\eta$ directly as opposed to approximating with $\tau_{\alpha}$ leads to vastly different interpretations of the SE relation.

Before examining these relations, we first consider the temperature-dependence of the viscosity, shown in \cref{fig:T-dependence}\textit{C}.
At the lowest $p_0$ studied, the variation in $\eta$ is consistent with the expectation for a standard glassformer, showing a sharp divergence over a narrow range of temperatures.
We note that in determining $\eta$ in the low stress regime, rather than observing Newtonian flow for all model parameters, for $p_0\leq3.825$ the model displays shear thinning at low-$T$.
These state points coincide with regions of the model that display hexatic ordering in the absence of an applied shear.
In these cases, the reported $\eta$ value is for the lowest applied stress that resulted in a measurable $\eta$ value and the true viscosity at the yield point may be higher.
For $p_0=3.85$, we observe that $\eta$ varies by less than two orders of magnitude over a range of $T$ for which $\tau_{\alpha}$ varies by more than four decades. 
This surprisingly small variation further highlights the decoupling between $\eta$ and $\tau_\alpha$ that can be found in this model.

\begin{figure*}[t]
\centering
\includegraphics[width=0.9\textwidth]{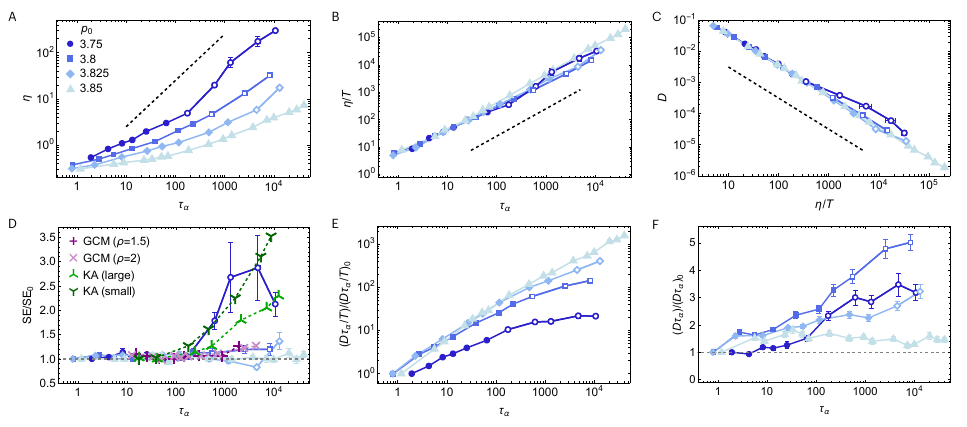}
\caption{
(\textit{A}) Dependence of the viscosity $\eta$ on the alpha-relaxation time $\tau_{\alpha}$ for different values of the shape index $p_0$. The color scale for $p_0$ values is the same across all figure panels and open markers indicate points for which hexatic ordering is observed in the structure.
(\textit{B}) Dependence of the ratio $\eta/T$ on $\tau_{\alpha}$. Dashed black lines in panels \textit{A} and \textit{B} indicate slope 1.
(\textit{C}) Relationship between the diffusivity $D$ and the ratio $\eta/T$ for different values of $p_0$.
The dashed black line indicates slope -1, which is expected when the Stokes-Einstein-Sutherland (SE) relation holds.
(\textit{D}) Comparison between values of the SE ratio $\text{SE}=D\eta/T$, scaled by the value $\text{SE}_0$ measured at the highest $T$ for each $p_0$, for the Voronoi model (blue points), the Gaussian Core Model (GCM) at two densities $\rho$ (purple points) and the large and small particles in the Kob-Andersen (KA) model (green points). GCM and KA data obtained from Ref.~\cite{Ikeda2011PRL}. 
(\textit{E}) Using $\tau_{\alpha}$ instead of $\eta$ in the SE ratio (note the logarithmic axis scale), or 
(\textit{F}) replacing $\eta/T$ with $\tau_{\alpha}$ would indicate stronger SE violations than are observed when the direct viscosity measurement is used (panel \textit{C}). 
Error bars in \textit{A--C}, which are typically smaller than the marker size, indicate the standard error across ten configurations while in \textit{D--F} they represent the propagated uncertainties.
}
\label{fig:SE}
\end{figure*}

Returning to the relations between $\eta$ and $\tau_{\alpha}$, \cref{fig:SE}\textit{A, B} show how $\eta$ and $\eta/T$ vary with $\tau_{\alpha}$ for different $p_0$ values. 
We observe that $\eta$ exhibits sublinear $p_0$-dependent scaling with $\tau_{\alpha}$, indicating that $\tau_{\alpha}\propto\eta$ is not a good approximation in this system.
By contrast, the scaling of $\eta/T$ with $\tau_{\alpha}$ shows minimal dependence on $p_0$ and $\tau_{\alpha}\propto\eta/T$ appears to be a reasonable approximation. 
Given the unusual scaling of both $\tau_{\alpha}$ and $\eta$ with temperature, it is somewhat surprising that this proportionality appears to hold reasonably well.
However, as we will show in the following section, using these approximations has important consequences for behavior of the SE relation.\\

\textbf{Stokes-Einstein-Sutherland relation.}
The breakdown of the SE relation as a system is cooled is often considered a hallmark of glassy materials~\cite{Ediger2000} and is taken as an indirect signature of the formation of DH.
We investigate this relation for the Voronoi model by examining the relation between $D$ and $\eta/T$ for different $p_0$ values, as shown in \cref{fig:SE}\textit{C}.
Consistent with the idea that at low $p_0$ these models cross over to normal glassy behavior, we observe that for $p_0=3.75$, the data shows the expected deviation from the SE relation at lower $T$.
However, for the higher $p_0$ values this deviation is not present, and the SE relation appears to be obeyed across the range of temperatures studied. 
To further quantify any deviations from the SE relation, we plot $\text{SE}/\text{SE}_0$ in \cref{fig:SE}\textit{D}, where $\text{SE}=D\eta/T$ and $\text{SE}_0$ is the value of $\text{SE}$ at the highest $T$ studied for each $p_0$.
This indicates a slight SE violation for $p_0=3.80, 3.825$ at the lowest temperatures (highest $\tau_{\alpha}$ values), but still shows no systematic deviation from the SE relation at the highest $p_0$ value.
These results suggest that DH may be suppressed within the model, which we explore further in the following section.

\begin{figure*}
\centering
\includegraphics[width=0.7\textwidth]{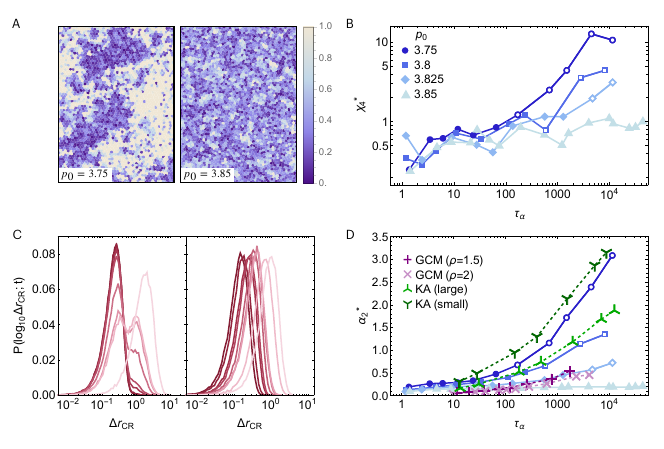}
\caption{
(\textit{A}) Cage-relative displacement field in a region of the system after time $t\sim\tau_{\alpha}$ has elapsed in a fully thermalized configuration for $p_0=3.75$ and $3.85$ with $\tau_{\alpha}\sim10^4$. Cells are plotted at their initial positions and colored according to their displacement at $t\sim\tau_{\alpha}$.
(\textit{B}) Peak height of $\chi_4(t)$ 
(\textit{C}) Probability distribution of $\log_{10}$ of the CR displacement field for $p_0=3.75$ (left) $p_0=3.85$ (right) for $\tau_{\alpha}\sim 10^4$. Individual curves show the distribution at different times, ranging from $t \ll\tau_{\alpha}$ (dark red) to $t \gg \tau_{\alpha}$ (light pink).
(\textit{D}) Peak value of the distribution $\alpha_2(t)$ as a function of $\tau_{\alpha}$ for different $p_0$ values in the Voronoi model. Also shown are data points for the GCM and KA model, taken from Ref.~\cite{Ikeda2011PRL}.
}
\label{fig:dh}
\end{figure*}

To compare these results to those of other model glassformers, \cref{fig:SE}\textit{D} also shows data for the Kob-Andersen (KA) model~\cite{Kob1997} and the Gaussian Core Model (GCM)~\cite{Stillinger1976,Stillinger1979PRB,Stillinger1979JChemPhys}, reproduced from Ref.~\cite{Ikeda2011PRL}.
The KA model is a frequently studied fragile glassformer consisting of a binary mixture of particles of different radius that interact through a Lennard-Jones potential.
This model shows the expected SE breakdown, comparable to the results for $p_0=3.75$.
Meanwhile, the GCM describes an ultrasoft glassformer where particles interact with each other via a Gaussian repulsion.
It displays weak SE breakdown and has DH that are different from those of typical glassformers~\cite{Ikeda2011PRL,Ikeda2011JChemPhys,Coslovich2016}.
In the GCM, the SE breakdown begins at later timescales and the growth of the SE ratio by the largest $\tau_{\alpha}$ studied is significantly smaller than for the KA model.
We observe that the results for the Voronoi model for $p_0=3.825$ are comparable to the GCM at both densities plotted while the ratio for $p_0=3.85$ is notably smaller than in either the GCM or KA model.

We emphasize that our decision to determine the viscosity directly in this system, rather than using $\tau_{\alpha}$ as a proxy for either $\eta$ or $\eta/T$ has a profound impact on our conclusions. 
Using $D\tau_{\alpha}/T$ for the SE relation suggests that the SE ratio increases by several orders of magnitude across the range of $\tau_{\alpha}$ studied, with the strongest violation at the highest $p_0$ value, as shown in \cref{fig:SE}\textit{E}. 
By contrast, using $D\tau_{\alpha}$  (\cref{fig:SE}\textit{F}) suggests SE violations more in line with those observed in other glassformers~\cite{Shi2013,Bouhadja2014,Ozawa2016}, but still much stronger SE violations for $p_0\geq 3.8$ than are observed when using the direct $\eta$ measurement.
The discrepancies between these approximations for the SE ratio and the value obtained using direct measurements of $\eta$ highlight the importance of determining $\eta$ when examining SE breakdown, particularly in materials with anomalous glassy dynamics.\\

\textbf{Dynamical heterogeneities.}
A natural hypothesis from the above -- given that SE violations are typically understood to be a consequence of dynamical heterogeneity (DH)~\cite{Ediger2000,Berthier2011} -- is that this model may be surprisingly dynamically \emph{homogeneous}.
To test this hypothesis, we consider several interrelated measures of DH which each reveal different features of the statistics of spatially correlated dynamics.
In models with atypical glassy dynamics, these different measures can lead to different conclusions about the nature of the DH~\cite{Ikeda2011JChemPhys,Coslovich2016,Nandi2021}.
We first directly plot the cage-relative (CR) displacement field for samples with $p_0=3.75$ and $3.85$ after time $t\sim\tau_{\alpha}$ has passed since some reference time with $\tau_{\alpha}\sim 10^4$. 
The results, shown in \cref{fig:dh}\textit{A}, indicate regions of high and low mobility for $p_0=3.75$, suggesting the presence of DH. 
By contrast, for $p_0=3.85$ there is no clear separation of regions with higher or lower displacements, suggesting more homogeneous dynamics.

We next consider the peak height $\chi_4^*$ of the four-point dynamic susceptibility $\chi_4(t)$ (see Methods).
In a typical glassformer, $\chi_4^*$ is expected to grow as the temperature decreases, and this is often interpreted to quantify a characteristic length scale related to the size of regions undergoing collective rearrangement~\cite{Berthier2011}.
As shown in \cref{fig:dh}\textit{B}, we observe that $\chi_4^*$ grows slowly with $\tau_{\alpha}$ for $p_0=3.75$, consistent with previous reports for $\chi_4(t)$ in this model~\cite{Sussman2018EPL}.
For $p_0=3.825$, $3.85$ there is slight growth in the peak height at the highest $\tau_{\alpha}$ values.
However, for $p_0=3.85$, the peak height shows only a very slight increase with $\tau_{\alpha}$, providing a quantitative indicator of a lack of correlated dynamics in the model at higher $p_0$ values.

Another, quite direct indicator of the presence of DH can be obtained through the probability distribution $P(\log_{10}\Delta r_{\text{CR}}; t)$, which is proportional to the van Hove function $G_s(\Delta r_{\text{CR}}; t)$.
Here $\Delta r_{\text{CR}}$ is the CR displacement after time $t$ has elapsed.
This distribution is shown in \cref{fig:dh}\textit{C} for $p_0=3.75$ (left) and $p_0=3.85$ (right) for temperatures that give $\tau_{\alpha}\sim 10^4$ at a range of $t$ values.
In a typical glassformer, this distribution becomes bimodal for time scales $t\sim \tau_{\alpha}$ due to the separation of fast and slow dynamics~\cite{Flenner2005,Ikeda2011JChemPhys,Coslovich2016}, which is what we observe for $p_0=3.75$.
By contrast, for $p_0=3.85$ we observe the distribution has a single peak that shifts to larger displacements at longer time scales while maintaining a constant height.
This trend is consistent with that expected for a Gaussian process~\cite{Ikeda2011JChemPhys}, indicating suppressed DH at higher $p_0$ values.
Note that while here the measurements of $\chi_4(t)$ and $P(\log_{10}\Delta r_{\text{CR}}; t)$ both indicate suppressed DH, these two measures do not always point to the same conclusion, as we discuss further in the following section.

A final quantity that can also be used to indicate the presence of DH, and can be compared to other models, is the non-Gaussian parameter $\alpha_2(t)$ defined in 2d by~\cite{Rahman1964, Huang2015}
\begin{equation}
\alpha_2(t) = \frac{1}{2}\frac{\left<\Delta r_{\text{CR}}^4(t)\right>}{\left<\Delta r_{\text{CR}}^2(t)\right>^2}-1.
\label{eq:alpha2}
\end{equation}
This quantity is zero if the distribution of particle displacements is Gaussian and nonzero otherwise, with a typical glassformer displaying a single-peaked distribution that grows in height as the temperature is lowered.
\Cref{fig:dh}\textit{D} shows the peak height $\alpha_2^*$ as a function of $\tau_{\alpha}$ for each $p_0$, which shows that the peak height decreases with $p_0$ for a given $\tau_{\alpha}$, providing further evidence of a reduction in DH as $p_0$ is increased.
Comparing these values to the KA model and GCM using data from Ref.~\cite{Ikeda2011PRL}, we observe that the trends are consistent to those observed for the SE ratio in \cref{fig:SE}\textit{D}, indicating correlation between the degree of DH and level of SE breakdown.

\begin{figure}
\centering
\includegraphics[width=0.5\textwidth]{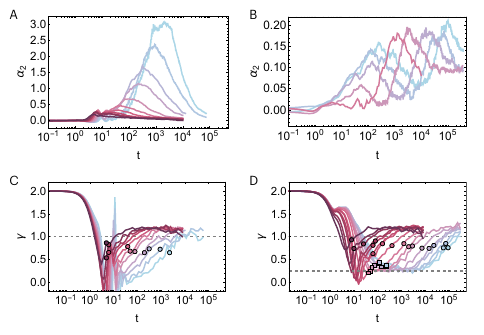}
\caption{
(\textit{A}) Non-Gaussian parameter $\alpha_2(t)$ for $p_0=3.75$ for different temperatures. 
(\textit{B}) $\alpha_2(t)$ for $p_0=3.85$ for five temperatures in the $T\leq 0.0005$ regime. 
(\textit{C, D}) Time-dependence of the exponent $\gamma(t)$ of the CR MSD for
(\textit{C}) $p_0=3.75$ and (\textit{D}) $p_0=3.85$ for each $T$. Plotted circles indicate the time at which the primary peak in $\alpha_2$ occurs for each $T$. Square markers for $p_0=3.85$ mark the time of the secondary peak.
Color scales in all panels indicate the temperature, ranging from low (light blue) to high (dark red).
}
\label{fig:alpha2}
\end{figure}

Looking more closely at the behavior of $\alpha_2(t)$, we observe that the $p_0=3.85$ data is qualitatively different to the lower $p_0$ data.
For $p_0\leq 3.825$, we observe the expected single peak in $\alpha_2(t)$, as shown in \cref{fig:alpha2}\textit{A} for $p_0=3.75$. 
However, for $p_0=3.85$ we observe a secondary peak at time scales much shorter than $\tau_{\alpha}$ for $T\lesssim 0.005$, as shown in \cref{fig:alpha2}\textit{B} for select temperatures in this lower $T$ regime.
The presence of this additional peak in $\alpha_2(t)$ for $p_0=3.85$ suggests the presence of an additional dynamical regime at lower $T$.
To our knowledge the only previously reported observation of multiple peaks in the non-Gaussian parameter are in a recently reported model aiming to bridge the gap between 3d and mean-field glasses by introducing ``pseudo-neighbor'' interactions for each particle~\cite{Nandi2021}.
In Ref.~\cite{Nandi2021}, the additional peak was observed in the regimes of the model that are closest to mean-field and the height of this peak increased as the model became more mean-field-like.
In contrast to the Voronoi model, in that model the additional peaks were only present at \emph{higher} temperatures and appeared at longer time-scales than the primary peak in $\alpha_2(t)$.

To explore the dynamics associated with this additional peak in $\alpha_2(t)$, we examine the time dependence of the exponent $\gamma(t)$ of the CR MSD (see Methods), defined by $\expval{\Delta^{CR} (t)}\propto t^{\gamma(t)}$.
At short time scales, the motion displays the expected ballistic ($\gamma(t)=2$) behavior, shown in \cref{fig:alpha2}\textit{C, D} for $p_0=3.75$ and $3.85$ respectively, while at long time scales the motion is super-diffusive ($\gamma(t)\sim 1.2$) up to the time scale studied.
This is likely due to caging effects in the system, and we expect that at long enough time frames the motion will become diffusive~\cite{Weeks2002,Flenner2019}.
At intermediate time scales, the motion is sub-diffusive, as expected due to caging effects~\cite{Weeks2002,Shiba2018}, resulting in a plateau in the CR MSD.
The location of the primary peak of $\alpha_2(t)$ (circle markers in \cref{fig:alpha2}\textit{C, D}) occurs as the system leaves the sub-diffusive regime, consistent with the observations in Ref.~\cite{Nandi2021}.
The secondary peaks for $p_0=3.85$ (square markers in \cref{fig:alpha2}\textit{D}), where present, appear to coincide with the start of an extended plateau with $\gamma(t)\sim 0.25$ that ends at times corresponding to the local minimum of $\alpha_2(t)$  between the two peaks.
More work is needed to establish how robust this subdiffusive regime is, but we note that $\expval{\Delta(t)} \propto t^{1/4}$ is observed in the monomer displacements in entangled linear polymers.
There the scaling arises from monomers constrained to diffuse along a reduced-dimensional region of space (there, the primitive path)~\cite{Rubinstein2003}, and we speculate that the unusual geometry of the zero-energy manifold of the large-$p_0$ Voronoi model~\cite{Pinto2022} may give rise to a similar effect.

\section*{Discussion}
Our results demonstrate that the dynamics of the 2d Voronoi model are highly unusual compared to those of standard glassformers and that the nature of the dynamics can be tuned by varying the shape index $p_0$.
This goes beyond previous reports focused on the unusual scaling of $\tau_\alpha$ with temperature~\cite{Sussman2018EPL,Li2021}.
In the low-$p_0$ regime, the model displays SE breakdown and the corresponding formation of DH expected as the dynamics slows down.
However, as $p_0$ is increased, the slowing down is not accompanied by breakdown of the SE relation and the system remains dynamically homogeneous.
At the highest $p_0$ studied, these dynamics are accompanied by an additional dynamical regime, characterized by a plateau in the slope of the CR MSD and a corresponding additional peak in $\alpha_2(t)$.
These findings suggest that this model may be able to tune between different universality classes governing disordered dynamics: a standard glassforming one and a potentially new universality class, possibly with a zero-temperature critical point as is possible in other 2d systems~\cite{Berthier2019}, which is characterized by a suite of anomalous low-temperature behavior.
Further work is required to explore this transition and determine the relevant critical exponents.

The dynamics of the Voronoi model displays several similarities with 3d models that emulate mean-field type behavior~\cite{Ikeda2011PRL,Ikeda2011JChemPhys,Coslovich2016,Mari2011,Nandi2021}. 
Both the GCM~\cite{Ikeda2011PRL,Ikeda2011JChemPhys,Coslovich2016} and models that incorporate additional interactions to approach the mean-field limit from 3d~\cite{Mari2011,Nandi2021} display the near-Gaussian statistics of particle displacements observed here at higher $p_0$.
However, measuring $\chi_4(t)$ reveals notable differences between these models.
In the GCM, $\chi_4^*$ grows strongly as the temperature is decreased, signaling the presence of giant dynamical fluctuations in the model despite the near-Gaussian statistics of the displacements~\cite{Coslovich2016}.
By contrast, in other mean-field approaching models~\cite{Mari2011,Nandi2021}, the peak heights decrease as the system becomes more mean-field-like, consistent with our observations in the 2d Voronoi model.
While the exact nature of the information conveyed by quantities such as $\chi_4(t)$ and $\alpha_2(t)$, and their relation to each other, still needs further exploration~\cite{Nandi2021}, our results paint a uniform picture of suppressed DH in the Voronoi model at higher $p_0$ values, with dynamics that are closer to the behavior of some mean-field-like models than to standard glassformers.

The relation between fragility  -- related to the temperature dependence of the viscosity of a system close to the glass transition~\cite{Debenedetti2001} -- and the resulting amount of SE breakdown and strength of DH has been a topic of much debate.
The fragility is negatively correlated with the stretching exponent $\beta$ characterizing the decay of the self-intermediate scattering function~\cite{Bohmer1993,Niss2007,Xia2001}, which itself is correlated with the degree of DH present.
As such, it is often assumed that more fragile materials will display stronger SE breakdown, mostly based on data in 3d.
While systems such as those studied in Refs.~\cite{Bouhadja2014,Ozawa2016} do show this correlation, several other systems do not~\cite{Chen2006,Jung2004,Nandi2021,Sengupta2013,Nandi2021,Dyre2007}.
The relation between fragility and level of SE breakdown may depend on the spatial dimension $d$, with studies suggesting the amount of SE breakdown decreases as $d$ increases~\cite{Sengupta2013,Nandi2021,Adhikari2021}. 
Whether this is accompanied by an increase~\cite{Sengupta2013,Nandi2021} or decrease~\cite{Adhikari2021} in the fragility is again inconclusive.
If there is a correlation between the fragility and degree of DH, this would predict that the ultra-strong regime of the 2d Voronoi model~\cite{Sussman2018EPL,Li2018}, would have very little SE violation, with higher $p_0$ having less DH.
This is consistent with our observations of the SE breakdown and DH presented here.

Further theoretical insight into the DH may be gained through inhomogeneous mode coupling theory (MCT) analysis~\cite{Biroli2006} of the model, building upon recent MCT results that may be capturing the ultra-strong behavior of the vertex model~\cite{Pandey2024}.
Moreover, recent work on the dynamics of active glasses has shown that details of the DH can be very different between equilibrium and active variants of standard model glassformers, despite similar relaxation dynamics between the two model variants~\cite{Paul2023}.
As such, studying the DH in an active Voronoi model, such as the self-propelled Voronoi model~\cite{Bi2016}, would give insight into which anomalous features of the DH are conserved across different microscopic dynamics.

We finally note that experiments on real cell monolayers have reported the presence of DH based on measurements of the velocity field~\cite{Angelini2011} and $\chi_4(t)$~\cite{Malinverno2017,Vishwakarma2020,Cerbino2021}.
The strength of the DH has been shown to be correlated with the level of coordinated motion and forces in the monolayer~\cite{Vishwakarma2020}, and also to decrease over time~\cite{Malinverno2017,Cerbino2021}.
A complicating feature, though, is the presence of large fluctuations in cell positions and shape that often occur with comparatively few neighbor-exchange events~\cite{Angelini2011}.
Thus, to draw a meaningful comparison between experimental results and the results presented here would require undertaking CR measurements on the real cells, separating out DH due to cell rearrangements and those due to cell-cage-scale fluctuations.
Exploring this relationship, and the implications of our results for understanding fundamental similarities and differences between the physics underlying different cell models remain exciting directions for future work.


\section*{Methods}
\textbf{Alpha-relaxation time $\vb*{\tau_{\alpha}}$.}
We estimate $\tau_{\alpha}$ from the decay of the cage-relative self-intermediate scattering function (CR SISF), which we use to remove the effect of long-wavelength Mermin-Wagner fluctuations present in 2d systems~\cite{Mazoyer2009,Vivek2017,Illing2017,Shiba2019}.
This is given by
\begin{equation}
F_{s}^{CR}(q,t) = \frac{1}{N} \sum_{i=1}^N \expval{\mathrm{e}^{i\vb{q}\vdot[\Delta\vb{r}_i(t)-\Delta\vb{r}_i^{\text{cage}}(t)]}},
\end{equation}
where $\Delta\vb{r}_i(t)=\vb{r}_i(t)-\vb{r}_i(0)$, $\Delta\vb{r}_i^{\text{cage}} = \frac{1}{N_i}\sum_{j=1}^{N_i}[\vb{r}_j(t)-\vb{r}_j(0)]$  gives the displacement of the cage formed by the $N_i$ neighbors of particle $i$ at $t=0$, and $q = \abs{\vb{q}}$ is the position of the peak in the static structure factor.
The value of $\tau_{\alpha}$ is estimated by fitting a stretched exponential of the form $A\exp(-t/\tau_{\alpha})^{\beta}$ to the long-time tail of $F_s(q,t)$, where $\tau_{\alpha}$, $A$ and $\beta$ are fitting parameters.
The final value for each $\tau_{\alpha}$ is then obtained by averaging the fit values across ten configurations for each state point.
We investigate a range of temperatures at each $p_0$ that correspond to $1\lesssim\tau_{\alpha}\lesssim10^4$.\\

\textbf{Self-diffusion constant $\vb{D}$.}
We obtain the self-diffusion constant $D$ from the mean-squared displacement (MSD), $\Delta(t) = \expval{\Delta r^2(t)}$, from which we expect that at long times 
\begin{equation}
    \Delta (t) \sim 2 d D t,
\end{equation}
with $d$ being the dimension of the system. 
We calculate the MSD at time $t=t_{\text{max}}$, with $t_{\text{max}}\geq 10\tau_{\alpha}$ being the duration of the simulation after an initial thermalization period (which is itself of order $10\tau_{\alpha}$), and use this to estimate $D$ for a single configuration. 
The final $D$ estimate is determined from the mean value from ten configurations at each state point.
We choose to use the standard MSD rather than the CR MSD because the CR MSD does not reach the diffusive regime within the timeframe of our simulations (as discussed in Results), whereas the standard MSD does.
At long enough time scales we expect the CR MSD to asymptotically approach the MSD~\cite{Flenner2019}, leading to the same estimate for $D$.\\

\textbf{Viscosity $\vb*{\eta}$.}
We measure the viscosity $\eta$ using reverse nonequilibrium molecular dynamics (RNEMD)~\cite{MullerPlathe1997,MullerPlathe1999, Kuang2010,Kuang2012}, in which an unphysical momentum flux $j_y(p_x)$ is imposed on the system in the $x$-direction, inducing a gradient in the $x$-velocity $v_x$ along the $y$-direction.
The system is divided into horizontal slabs with a forward momentum flux imposed on the central slab and a backward momentum flux imposed on the bottom slab.
The fluxes are chosen to ensure that the linear momentum and kinetic energy are conserved throughout the system.
The viscosity is then $\eta=-j_y(p_x)/\pdv{v_x}{y}$.
We implement the velocity shearing and scaling (VSS) RNEMD method~\cite{Kuang2012}, 
which reduces perturbations compared to earlier RNEMD methods by distributing the momentum flux across all particles in the manipulated slab.
We set the momentum swap frequency to $f=5\dd t$ and have verified that the results are consistent across a range of choices of $f$.
To allow the velocity profile to develop, we run simulations for $50\,000\tau$ from a thermalized configuration before determining $\eta$ by averaging $\pdv{v_x}{y}$ over a duration of at least $60\,000\tau$.\\

We determine $\eta$ for each $p_0$ and $T$ by sampling at least four fluxes in the range $10^{-5}\lesssim j_y(p_x) \lesssim 10^{-3}$ with the value of $\eta$ at each state point averaged across ten configurations.
For state points where there is no hexatic ordering present, we observe that the flow is Newtonian and determine the final estimate for $\eta$ by averaging its value across the range of fluxes studied.
When shear thinning is present, $\eta$ is obtained from the lowest $j_y(p_x)$ that yields a value, averaged across ten configurations.\\

\textbf{Four-point susceptibility $\vb*{\chi_4(t)}$.} 
The four-point dynamic susceptibility $\chi_4(t)$ is defined by
\begin{equation}
\chi_4(t) = N \expval{\delta F_s^{\text{CR}}(\vb{q},t)\delta F_s^{\text{CR}}(\vb{q},t)},
\end{equation}
where $\delta F_s^{\text{CR}}(\vb{q},t)$ is the difference between the instantaneous CR SISF and its mean, and the value of $\vb{q}$ used corresponds to the maximum of the static structure factor~\cite{Szamel2006}.\\

\textbf{CR MSD $\expval{\vb*{\Delta^{CR} (t)}}$.}
The cage-relative mean-squared displacement (CR MSD) is defined by~\cite{Mazoyer2009}
\begin{equation}
   \expval{\Delta^{CR} (t)} = \frac{1}{N}\expval{\sum_{i=1}^N\left[\Delta\vb{r}_i(t)-\Delta\vb{r}_i^{\text{cage}}(t)\right]^2},
\end{equation}
where $\expval{\quad}$ denotes averaging across configurations.
Results for the CR MSD are averaged across ten configurations at each $T$ and $p_0$.

\section*{Acknowledgements}
We thank Tomilola Obadiya and Eric Weeks for helpful conversations.
HSA acknowledges funding from the Tarbutton Postdoctoral Fellowship.
This material is based upon work supported by the National Science Foundation under Grant No. DMR-2143815.
This research used the Delta advanced computing and data resource which is supported by the National Science Foundation (award OAC 2005572) and the State of Illinois. Delta is a joint effort of the University of Illinois Urbana-Champaign and its National Center for Supercomputing Applications.

\section*{Author contributions}
H.S.A. and D.M.S. designed research;
H.S.A. and C.L. performed research;
H.S.A., C.L. and D.M.S. analyzed data;
H.S.A. and D.M.S. wrote the paper.

\bibliography{stokesEinsteinVoronoi_sources}

\end{document}